# Channel Coding and Decoding in a Relay System Operated with Physical-layer Network Coding

Shengli Zhang, Soung Chang Liew

Department of Information Engineering
The Chinese University of Hong Kong
New Territories, Hong Kong

**Abstract:**

This paper investigates link-by-link channel-coded PNC (Physical layer Network Coding), in which a critical process at the relay is to transform the superimposed channel-coded packets received from the two end nodes (plus noise), $Y_3 = X_1 + X_2 + W_3$, to the network-coded combination of the source packets, $S_1 \oplus S_2$. This is in contrast to the traditional multiple-access problem, in which the goal is to obtain both $S_1$ and $S_2$ explicitly at the relay node. Trying to obtain $S_1$ and $S_2$ explicitly is an overkill if we are only interested in $S_1 \oplus S_2$. In this paper, we refer to the transformation $Y_3 \to S_1 \oplus S_2$ as the Channel-decoding-Network-Coding process (CNC) in that it involves both channel decoding and network coding operations. This paper shows that if we adopt the Repeat Accumulate (RA) channel code at the two end nodes, then there is a compatible decoder at the relay that can perform the transformation $Y_3 \to S_1 \oplus S_2$ efficiently. Specifically, we redesign the belief propagation decoding algorithm of the RA code for traditional point-to-point channel to suit the need of the PNC multiple-access channel. Simulation results show that our new scheme outperforms the previously proposed schemes significantly in terms of BER without added complexity.

## I. Introduction

The two-way relay channel (TWRC) is a fundamental network structure of much interest to the wireless communications research community. Application of network coding in TWRC, in particular, has attracted intense interest recently. The first proposal of network coding for TWRC can be traced to [1], in which network coding is applied at the relay node to exploit the broadcast nature of the wireless medium. With respect to Fig. 1, the scheme works as follows. Node $N_1$ sends node $N_3$ its packet. Through another orthogonal channel, node $N_2$ sends node $N_3$ its packet. Then $N_3$ mixes the information of $N_1$ and $N_2$ to form a network-coded packet and broadcasts it to $N_1$ and $N_2$. In this way, the number of time slots needed to exchange one packet is three. The scheme in [1] regards network coding as an upper layer technique, and separates it from other lower-layer processes such as modulation and channel coding. In [2, 3], this scheme was further extended to combine with channel coding.

In [4], we proposed a new network coding scheme called Physical-layer Network Coding (PNC). PNC was originally inspired by the observation that the relay node $N_3$ does not need to know the individual contents of the source packets, $S_1$ and $S_2$, to form the network-coded packet $S_1 \oplus S_2$, and that the needed information $S_1 \oplus S_2$ could be obtained even if the two end nodes were to transmit simultaneously to the relay in the same time slot. In particular, $N_3$ in PNC directly transforms the superimposed packets received to the network-coded packet $S_1 \oplus S_2$ for broadcast to $N_1$ and $N_2$. In this way, the number of time slots needed to exchange one packet is reduced from three to two with respect to the scheme in [1]. At the same time, the bit-error rate (BER) is also decreased [4].

An issue left open by [4] is the use of channel coding to achieve reliable communication. There are two ways to apply channel coding in PNC. First, channel coding could be applied on an end-by-end basis, in which only the end nodes, but not the relay node, perform channel encoding and decoding. We refer to this set-up as end-to-end coded PNC. Second, channel coding could be applied on a link-by-link basis, in which the end nodes as well as relay node perform channel encoding and decoding. In particular, the relay will first transform the superimposed channel-coded signals $Y_3 = X_1 + X_2 + W_3$ ($W_3$ is the noise at $N_3$) received from the end nodes to unchannel-coded but network-coded information $S_1 \oplus S_2$, and then channel-encode $S_1 \oplus S_2$ for broadcast to the end nodes. We refer to this set-up as link-by-link coded PNC. This paper investigates link-by-link coded

PNC schemes, focusing on the critical transformation process $Y_3 \to S_1 \oplus S_2$ therein. Note that the process of channel-encoding $S_1 \oplus S_2$ is the same as that for ordinary point-to-point channel, whereas the transformation $Y_3 \to S_1 \oplus S_2$ can be quite intricate and its implementation can affect the system performance significantly, as will be demonstrated in this paper. We refer to the process of $Y_3 \to S_1 \oplus S_2$ as the Channel-decoding-Network-Coding process (CNC).

Two straightforward link-by-link coded PNC schemes with different implementations of CNC can be found in the literature [5, 6]. Throughout this paper, lowercase letters will be used to denote symbols, and the corresponding uppercase letters will be used to denote packets containing the symbols. For example, $s_1$ denotes a source symbol from node $N_1$, while $S_1$ denotes an overall packet containing a sequence of source symbols. In the first scheme, the relay (i) explicitly decodes and extracts the two source packets $S_1$ and $S_2$ contained in the superimposed channel-coded packets $Y_3$ received from the end nodes; and (ii) combines the two source packets $S_1$ and $S_2$ to form the network-coded packet $S_1 \oplus S_2$. In the second scheme, the relay (i) maps each pair of superimposed channel-coded symbols $y_3$ contained in the overall superimposed packets $Y_3$ to an estimate of the network-coded symbol $x_1 \oplus x_2$ to form an interim packet $\widehat{X_1 \oplus X_2}$; and (ii) performs channel decoding on the interim packet $\widehat{X_1 \oplus X_2}$ to obtain the network-coded packet $S_1 \oplus S_2$.

The first scheme (in particular step (i) of it) falls under the framework of the generic multiple-access problem [7, Theorem 14.3.1]. To the best of our knowledge, the second scheme was first proposed and studied in [5, 6]. In [8, 9], the authors proved that the first and second schemes can approach the exchange capacity of TWRC in the low and high SNR regions, respectively, assuming all nodes use the same transmit power. In [10, 11], the results were extended to the case of different nodes using different transmit powers.

Two design principles for a good CNC scheme are as follows: (a) decoding of extraneous information not related to $S_1 \oplus S_2$ should be avoided so that unnecessary burdens are not imposed on the decoder; and (b) $X_1+X_2$ contains useful information for the decoding of $S_1 \oplus S_2$, and this useful information contained in $Y_3$ should contribute toward the decoding of $S_1 \oplus S_2$. Each of the above two schemes does not satisfy one of the principles. In particular, the first scheme does not make full use of the fact that it is not necessary for the relay to obtain the explicit individual source packets $S_1$ and $S_2$ from the end nodes, and the decoding of extraneous information $H(S_1, S_2 | S_1 \oplus S_2)$ in its step (i) results in unnecessary additional power requirements. For the second scheme, the PNC mapping in its step (i) discards useful information related to $S_1 \oplus S_2$ contained in $Y_3$. In other words, the two schemes underperform for the opposite reasons: the first scheme over-decodes, and the second scheme over-discards information.

This paper proposes a novel joint design of network coding and channel coding that attempts to adhere to the above design principles. In the new scheme, the relay (i) channel-decodes the superimposed channel-coded packets $Y_3$ to obtain the soft version of the arithmetic summation of the two source packets $S_1 + S_2$ (i.e., the PMF (probability mass function) of $S_1 + S_2$); (ii) transforms the superimposed source packets $S_1 + S_2$ (soft version) to the network-coded packet $S_1 \oplus S_2$. Compared with the first straightforward scheme, step (i) of the new scheme aims to obtain $S_1+S_2$, rather than individual $S_1$ and $S_2$ to reduce extraneous decoded information. In fact, if the channel decoder only aims to obtain $\Pr[s_1 + s_1 = 1]$ and $\Pr[s_1 + s_1 \neq 1]$ rather than the complete PMF covering $\Pr[s_1 + s_1 = 0]$, $\Pr[s_1 + s_1 = 1]$, and $\Pr[s_1 + s_1 = 2]$, then no extraneous information will be decoded, and the first design principle will be completely adhered to. Compared with the second straightforward scheme, in the channel-decoding process, step (i) of the new scheme directly processes on $Y_3$ while step (ii) of the second scheme processes decoding on $\widehat{X_1 \oplus X_2}$, where some information related to $S_1 \oplus S_2$ has already been lost.

Although the intuitive rationale for the new scheme is clear, it is not obvious that the special channel decoder needed for its step (i) exists. A main contribution of this paper is to provide the explicit construction of such a decoder based on the use of the Repeat Accumulate (RA) code [12,13]. Specifically, we redesign the belief propagation algorithm of the RA code for traditional point-to-point channel to suit the need of the PNC multiple-access channel. Simulation results show that our new scheme outperforms the previously proposed schemes significantly in terms of BER without added complexity in our decoder design.

The remainder of this paper is organized as follows. Section II presents our system model and provides formal definitions and classification of PNC. Section III puts forth the concept of our new link-by-link coded PNC scheme, while Section IV presents a specific design of the CNC decoder for it. We investigate the relative performance of CNC schemes in Section V. Section VI concludes this paper.

## II. System Model and Definitions

### A. System model

We consider the two-way relay channel as shown in Fig.1, in which nodes $N_1$ and $N_2$ exchange information with the help of relay node $N_3$. We assume that all nodes are half-duplex, i.e., a node cannot receive and transmit simultaneously. This is an assumption arising from practical considerations because it is difficult for the wireless nodes to remove the strong interference of its own transmitting signal from the received signal. We also assume that there is no direct link between node $N_1$ and $N_2$. An example in practice is a satellite communication system in which the two end nodes on the earth can only communicate with each other via the relay satellite.

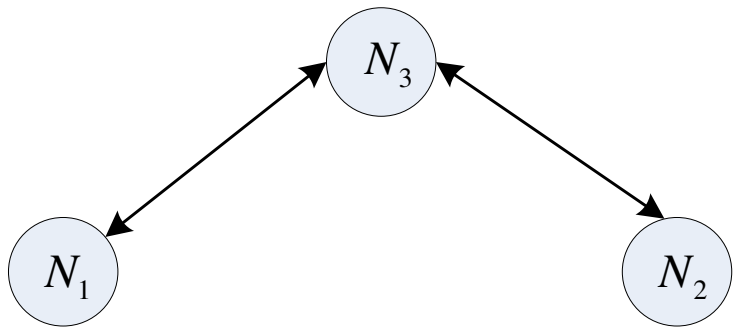

Fig 1:   Two-way relay channel.

In this paper, $S_i$ denotes the uncoded source packet of node $N_i$; $X_i$ denotes the corresponding packet after channel coding; $A_i$ denotes the corresponding transmitted packets after BPSK modulation; and $Y_i$ denotes the received base-band packet at node $N_i$. A lowercase letter, $s_i \in \{0,1\}$, $a_i \in \{-1,1\}$, $x_i \in \{0,1\}$, or $y_i \in \mathbb{R}$, denotes one symbol in the corresponding packet. We use $\Gamma_i$ to denote the channel coding scheme adopted by node $N_i$. Specifically,

$$X_i = \Gamma_i(S_i) \qquad S_i = \Gamma_i^{-1}(X_i) \tag{1}$$

We consider a two-phase transmission scheme consisting of an uplink phase and a downlink phase. In the uplink phase, $N_1$ and $N_2$ transmit to $N_3$ simultaneously. Therefore, $N_3$ receives

$$\begin{aligned}y_3' &= \sqrt{P_1}a_1 + \sqrt{P_2}a_2 + w_3' \\ &= \sqrt{P_1}(1-2x_1) + \sqrt{P_2}(1-2x_2) + w_3' \quad \text{s.t.} \quad P_1 + P_2 = 2\end{aligned} \tag{2}$$

where $w_3'$ is the noise at $N_3$, assumed to be Gaussian with variance $\sigma^2$ (identical for all the three nodes); and $P_i$ takes the transmit power and channel fading effect of $N_i$ into account. In (2), perfect synchronization is assumed. Synchronization is an important issue in PNC and other wireless communication systems. More details about it can be found in [5] and the references therein. With coherent soft decision demodulation, the received signal at $N_3$ can be expressed as

$$y_3 = -\left(y_3' - \sqrt{P_1} - \sqrt{P_2}\right) = \sqrt{4P_1}x_1 + \sqrt{4P_2}x_2 + w_3 \tag{3}$$

where the Gaussian noise $w_3 = -w_3' \in N(0,\sigma^2)$ and its vector version is $W_3$. Hereafter, we write the received packet $Y_3$ as a function of the transmitted packet $X_1+X_2$ without explicit explanation of the modulation-demodulation procedure.

In the downlink phase, $N_3$ generates a new packet $X_3$ based on the received packet $Y_3$, and broadcasts it to both $N_1$ and $N_2$. We can write the signals received by $N_1$ and $N_2$ as

$$y_1 = \sqrt{4P_3}x_3 + w_1 \qquad y_2 = \sqrt{4P_3}x_3 + w_2 \tag{4}$$

where, for simplicity, the channel gains for the channels from the relay node to $N_1$ and to $N_2$ are assumed to be the same. The target information $X_1$ ($X_2$) will be decoded from $Y_2$ ($Y_1$) at $N_2$ ($N_1$) with the help of its self-information. In general, $X_3$ must be a function of $Y_3$, which is in turn a function of $X_1$ and $X_2$. That is, $X_3 = f(Y_3)$ (note that $f$ may involve complex transformation and may not be a simple mapping). Part B below defines and classifies PNC.

*B. Definitions and classification of PNC*

***Definition 3.1 (PNC):*** *Physical-layer network coding is the coding operation which transforms the received baseband packet at $N_3$, $Y_3 = \sqrt{4P_1}X_1 + \sqrt{4P_2}X_2 + W_3$, to a network-coded packet $X_3 = f(Y_3)$ for relay, where $X_1$ and $X_2$ are the packets transmitted by $N_1$ and $N_2$ simultaneously to $N_3$.*

If the relay node does not perform any channel decoding and re-encoding operation (only the source node performs channel encoding and the sink node performs channel decoding), the PNC transformation in *Definition 3.1* then works in a symbol-by-symbol manner. The uppercase letters denoting packets could be replaced by lowercase letters denoting symbols in *Definition 3.1*. We refer to this as end-to-end coded PNC. Interested readers are referred to [14] for a study of end-to-end coded PNC.

By contrast, if channel coding is involved in the PNC transformation at the relay, each symbol in $X_3$ may depend on other symbols in $Y_3$ due to the correlation created by the channel coding. Therefore, the PNC transformation operates on a packet-by-packet basis, and the wireless uplinks and downlinks between the end nodes and the relay are separately protected by channel coding. We refer this set-up as link-by-link coded PNC. Because both $S_1$ and $S_2$ are assumed to be over GF(2) in this paper, we only consider network coding over GF(2) and hence the only nontrivial network coding operation is to form the modulo-2 sum (XOR) of the packet $S_1$ and $S_2$. And $X_3$ will be in the form of $\Gamma_3(S_1 \oplus S_2)$. The formal definition of link-by-link coded PNC is as follows:

***Definition 3.2 (Link-by-link Coded PNC):*** *Link-by-link coded PNC is the coding operation which transforms the received baseband packet at $N_3$, $Y_3 = \sqrt{4P_1}X_1 + \sqrt{4P_2}X_2 + W_3$, into a network-coded packet $X_3 = \Gamma_3(S_1 \oplus S_2) = \Gamma_3(h(Y_3))$ for relay, where $X_1$ and $X_2$ are the packets transmitted by $N_1$ and $N_2$ simultaneously to $N_3$.*

Unless stated otherwise, PNC hereafter means link-by-link coded PNC. Once $S_1 \oplus S_2$ is obtained, it is a straightforward process to channel-encode $S_1 \oplus S_2$ to obtain $\Gamma_3(S_1 \oplus S_2)$.

Therefore, key to PNC is the CNC process at the relay to obtain $S_1 \oplus S_2 = h(Y_3)$ from $Y_3$, defined as

***Definition 3.3 (CNC):*** *The Channel-decoding-Network-Coding process (CNC) is the process at the relay that transforms* $Y_3 = \sqrt{4P_1}X_1 + \sqrt{4P_2}X_2 + W_3$ *to* $S_1 \oplus S_2$.

Indeed, the study of this paper focuses on the CNC process, as the efficient implementation of it holds the key to the performance of a good link-by-link coded PNC system. For simplicity, we assume $P_1=P_2=1$ hereafter to focus on the basic idea of the proposed CNC. The discussion related to unequal power allocation (or where channel fading effects are taken into account) are given in the appendix.

### III. A Novel Link-by-link Coded PNC

In this section, we first briefly introduce two straightforward and well studied CNC schemes, CNC1 and CNC2. After that, we propose a new scheme, Arithmetic-sum CNC (ACNC), that performs the channel decoding specifically designed for network coding mapping at the relay node.

**CNC Design 1 (CNC1)**

In CNC1, the relay $N_3$ first decodes $S_1$ and $S_2$ from $Y_3$ separately. Note that this is in fact the well known multiple-access problem [7, Theorem 14.3.1]. With standard channel decoding, the relay can first decode one packet, say $S_1$, while regarding the other packet $S_2$ as interference, and can then decode $S_2$ after removing the decoded information $S_1$ from the received signal. Supposing SISO (soft input soft output) channel decoder is used, we can obtain the PMF (probability mass function) of the pair ($s_1$, $s_2$), denoted by $P_{s_1,s_2}(a,b) = \Pr(s_1 = a, s_2 = b | Y_3)$. Then, the relay node can directly combine them with network coding (XOR) to obtain $S_1 \oplus S_2$, as

$$s_1 \oplus s_2 = \begin{cases} 1 & \text{if } P_{s_1,s_2}(1,0) + P_{s_1,s_2}(0,1) \geq 0.5 \\ 0 & \text{else} \end{cases} \quad (5)$$

The block diagram of this scheme is shown in Fig. 2.

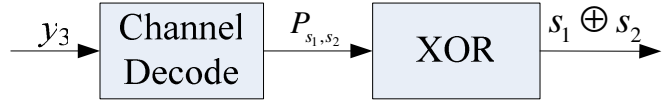

Fig 2: Block diagram of CNC1.

**CNC Design 2 (CNC2)**

In CNC2, the relay $N_3$ first estimates the PMF of $x_1 \oplus x_2$, denoted by $P_{x_1 \oplus x_2}(a) = \Pr(x_1 \oplus x_2 = a \mid y_3)$, from the received symbol $y_3$ with MMSE estimation (see [14] for details). Using the same linear channel codes at both end nodes (e.g., LDPC code is linear under binary addition, and the lattice code is linear under modulo addition [10, 15]), the packet $X_1 \oplus X_2$ is the codeword of $S_1 \oplus S_2$. By decoding the estimate of $X_1 \oplus X_2$, i.e. $P_{x_1 \oplus x_2}$, directly with a soft input decoder, the relay can obtain $S_1 \oplus S_2$. The block diagram of CNC2 is shown in Fig. 3.

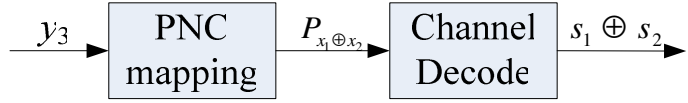

Fig 3: Block diagram of CNC2

CNC1 and CNC2 do not satisfy the two good-CNC design principles mentioned in the introduction. By decoding $S_1$ and $S_2$ explicitly, CNC1 obtains extraneous information unrelated to $S_1 \oplus S_2$, resulting in unnecessary power penalty. In CNC2, the PNC mapping from symbol $y_3$ to the PMF of $x_1 \oplus x_2$ discards useful information related to the decoding of the whole packet $S_1 \oplus S_2$. Our new scheme, Arithmetic-sum CNC design (ACNC), attempts to follow the two design principles.

**Arithmetic-sum CNC Design (ACNC)**

Our Arithmetic-sum CNC design, ACNC, works as follows. The relay first decodes $Y_3$ into to obtain the PMF of $s_1 + s_2$, denoted by $P_{s_1+s_2}(a) = \Pr(s_1 + s_2 | Y_3)$. Then, the relay could obtain the target information with the following PNC mapping:

$$s_1 \oplus s_2 = \begin{cases} 1 & \text{if } P_{s_1+s_2}(1) \geq 0.5 \\ 0 & \text{else} \end{cases}. \tag{6}$$

***Remark 1:*** From (6), we can see that the relay only needs to correctly decode the sign of $P_{s_1+s_2}(1) - P_{s_1+s_2}(2) - P_{s_1+s_2}(0)$. The individual probabilities of $P_{s_1+s_2}(1)$, $P_{s_1+s_2}(2)$ and $P_{s_1+s_2}(0)$ are not necessary.

The relay node finally encodes $S_1 \oplus S_2$ with standard channel encoder and broadcasts it to both end nodes. The block diagram of this scheme is shown in Fig. 4.

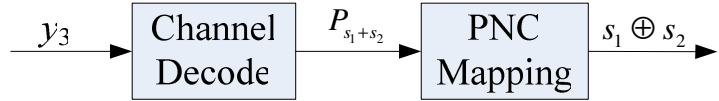

Fig 4: Block diagram of ACNC

We can see the advantages of ACNC as follows. First, in ACNC the relay directly decodes the received packet $Y_3$ to make full use of the information and dependency of symbols within the packet; by contrast, the symbol-level PNC mapping in CNC2 neglects the dependency among symbols created by the channel code. Second, in ACNC the channel decoder of the relay obtains the PMF of $s_1 + s_2$ which can be easily transformed to $s_1 \oplus s_2$ by symbol-level PNC mapping; by contrast, obtaining $s_1$ and $s_2$ explicitly as in CNC1 is unnecessary and such extraneous information constrains the reliable transmission rates of both $s_1$ and $s_2$.

The above intuition indicates that ACNC should perform best among the three link-by-link coded PNC schemes. In the Appendix of [19], we examine the three CNC schemes from an information-theoretic viewpoint. By assuming the existence of the special channel decoder needed in ACNC, and that it can reliably decode $S_1+S_2$ with a rate approaching the mutual information of the channel, we show that ACNC can substantially outperform both CNC1 and CNC2.

However, the special and practical channel decoder as needed in ACNC is completely new and has not been studied before. It is motivated by the special requirement of joint channel coding and physical layer network coding. In the next section, we propose a specific decoding algorithm for ACNC.

## IV. A Novel Channel Coding Scheme for ACNC

The analysis in the Appendix of [19] shows that CNC1 and CNC2 outperform non-PNC Straightforward Network Coding (SNC) significantly. However, there is still a significant gap between their performance and the theoretical upper bound. CNC1 underperforms in the high SNR region; CNC2 underperforms in the low SNR region; and they both underperform when SNR is in the vicinity of 0 dB. ACNC, on the other hand, has the potential to achieve good performance for all range of SNR. Motivated as such, this section proposes a new channel coding scheme for ACNC based on Repeat Accumulate (RA) code.

Although we focus on regular RA codes in this paper, extensions to other channel codes, such as LDPC codes and Turbo codes, are straightforward. RA codes were first proposed in [12]. They can be regarded as special LDPC codes whose decoding operation are of low complexity, or special Turbo codes whose encoding operation are of linear complexity. Despite its simple encoding and decoding structure, RA codes (especially some new versions of RA codes, such as IRA in [13]) can approach the Shannon capacity of the point-to-point channel.

We now introduce our novel channel decoding scheme in ACNC to perform the processing $Y_3 \to S_1 + S_2$ for an implementation of ACNC. The encoder at $N_1$ and $N_2$, and the decoder at the $N_3$ are as follows:

### A. Encoder at $N_1$ and $N_2$:

We assume $N_1$ and $N_2$ use the traditional encoder of RA codes. This means that the modification at the transmitter is not needed. The RA encoder has a very simple structure. As illustrated in Fig. 5, the input packet $S_i$ of the encoder is first repeated $q$ ($q \geq 3$) times. After that, the bits are interleaved and accumulated by binary summation $\oplus$ to generate the codeword $X_i$. We further assume that the interleave pattern and the repeat factor $q$ are the same for the two end nodes.

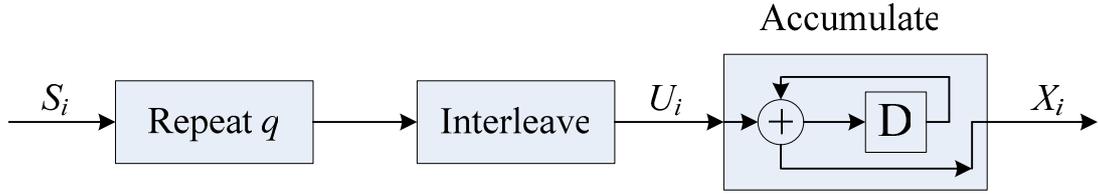

Fig 5:   Encoder of standard RA code.

*B. Decoder at $N_3$:*

The decoder at $N_3$ is different from the traditional RA decoder. This part provides the design of such a decoder along the following three steps: 1) construct a virtual encoder corresponding to the decoder; 2) construct the Tanner graph of the virtual code; 3) design the belief propagation algorithm based on the Tanner graph.

*Step 1*: *Virtual Encoder*

For ACNC, the decoder at relay can be regarded as processing the superposition of the two simultaneously received signals from $N_1$ and $N_2$ to generate the superposition of the two inputs of the encoders at $N_1$ and $N_2$. In the absence of noise, the received signals are the superposition of the two outputs of the encoders at $N_1$ and $N_2$. Thus, the decoding process at $N_3$ can be viewed as the inverse of the superposition of the encoding processes at $N_1$ and $N_2$. As such, the decoder at $N_3$ could conceptually be viewed as the decoder of a virtual encoder whose input $S_v$ and output $X_v$ are

$$S_v = S_1 + S_2 \qquad X_v = X_1 + X_2 \tag{7}$$

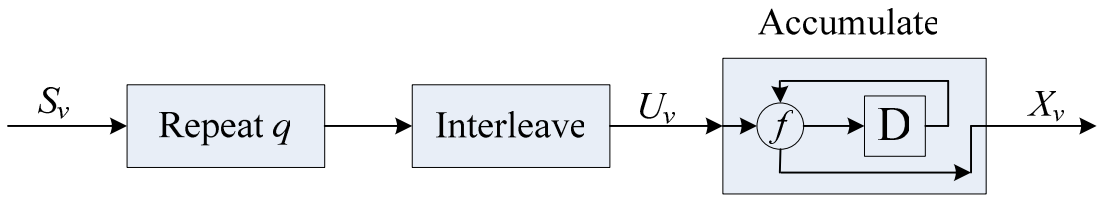

Fig 6:   The virtual encoder for ACNC.

The design of the decoder is intimately tied to the structure of this virtual encoder. As shown in Fig. 6, the virtual encoder has the same structure as the RA encoder in Fig. 5 except that the binary summation is now replaced by a general function $f$. Let us derive $f$ based on the specification in (7). Accordingly, the function $f$ in Fig. 6 needs to satisfy

$$x_v[k] = f(x_v[k-1], u_v[k]) = x_1[k] + x_2[k] \quad \text{when } s_v[j] = s_1[j] + s_2[j] \tag{8}$$

where $x_i[k]$ is the $k$-th coded bit of node $N_i$, $u_i[k]$ is the $k$-th interleaved bit of node $N_i$ and $s_i[j] = u_i[k]$ is the $j$-th information bit of $N_i$, and the index $j$ is determined by the interleaver, which is the same for both the end nodes' encoders and for the virtual encoder. Based on Fig. 5, the relations between $x_1[k]$, $x_2[k]$ and $s_1[j]$, $s_2[j]$ can be respectively expressed as

$$\begin{aligned} x_1[k] &= x_1[k-1] \oplus u_1[k] = x_1[k-1] \oplus s_1[j] \\ x_2[k] &= x_2[k-1] \oplus u_2[k] = x_2[k-1] \oplus s_2[j] \end{aligned} \tag{9}$$

Combining (8) and (9), we can obtain the expression of the function $f$ as

$$x_v[k] = f(x_v[k-1], u_v[k]) = x_1[k-1] \oplus s_1[j] + x_2[k-1] \oplus s_2[j]$$

$$= \begin{cases} 0 & \text{if } (x_1 + x_2)[k-1] = 2, (s_1 + s_2)[j] = 2 \\ 1 & \text{if } (x_1 + x_2)[k-1] = 2, (s_1 + s_2)[j] = 1 \\ 2 & \text{if } (x_1 + x_2)[k-1] = 2, (s_1 + s_2)[j] = 0 \\ 1 & \text{if } (x_1 + x_2)[k-1] = 1, (s_1 + s_2)[j] = 2 \\ 0 \text{ or } 2 & \text{if } (x_1 + x_2)[k-1] = 1, (s_1 + s_2)[j] = 1 \\ 1 & \text{if } (x_1 + x_2)[k-1] = 1, (s_1 + s_2)[j] = 0 \\ 2 & \text{if } (x_1 + x_2)[k-1] = 0, (s_1 + s_2)[j] = 2 \\ 1 & \text{if } (x_1 + x_2)[k-1] = 0, (s_1 + s_2)[j] = 1 \\ 0 & \text{if } (x_1 + x_2)[k-1] = 0, (s_1 + s_2)[j] = 0 \end{cases} \tag{10}$$

where $(s_1 + s_2)[i] = s_1[i] + s_2[i]$, $(x_1 + x_2)[i] = x_1[i] + x_2[i]$. It is easy to verify that the function $f$ in (10) satisfies the following two properties:

(a) $f(a, b) = f(b, a)$

(b) if $c = f(a, b)$, then $a = f(c, b)$, $b = f(c, a)$

for $a, b, c \in \{0, 1, 2\}$. The same properties are found in the traditional RA code where the accumulate function is XOR. Indeed, underlying the beauty of the RA encoding and decoding mechanisms are properties (a) and (b).

***Remark 2:*** From (10), we can see that the probability of $x_v[k]=2$ and the probability of $x_v[k]=0$ do not depend on the information sequence $S_1+S_2$ for any $k>0$ when given $x_v[k]\neq 1$ and $x_v[0]=1$. Due to the two symmetric properties (a) and (b) of *f*, the decoded symbol $s_1[k]+s_2[k]$ would also equal 0 or 2 in a random way when it does not equal 1. However, this is innocuous to the decoding of $s_1[k]\oplus s_2[k]$. Also, the fact that $s_1[k]+s_2[k]=0$ or 2 in a random way when $s_1[k]+s_2[k]\neq 1$ means that the decoder does not attempt to acquire any extraneous information.

*Step 2: Tanner Graph*

RA code can be described with the well known Tanner graph, which is the basis of the widely used belief propagation decoding algorithm [16]. Consider the Tanner graph of the virtual RA code in Fig. 7, which is constructed based on the encoder in Fig. 6. In Fig. 7, an information node, a vertex belonging to *S*, corresponds to an input bit; and a code node, a vertex belonging to *X*, corresponds to an output bit of the encoder. The information and code nodes are referred to as the variable nodes. An evidence node, a vertex belonging to *Y*, corresponds to a received symbol in $Y_3$. In Tanner graph, a check node, a vertex belonging to *C*, represents a "local constraint" on a subset of variable nodes, i.e., the values of the variable nodes connected to a check node should satisfy a predefined equation. For example, the value of any one of the three variable nodes connected to one check node in Fig. 7 should be the output of the *f* function with the values of the other two variable nodes as inputs.

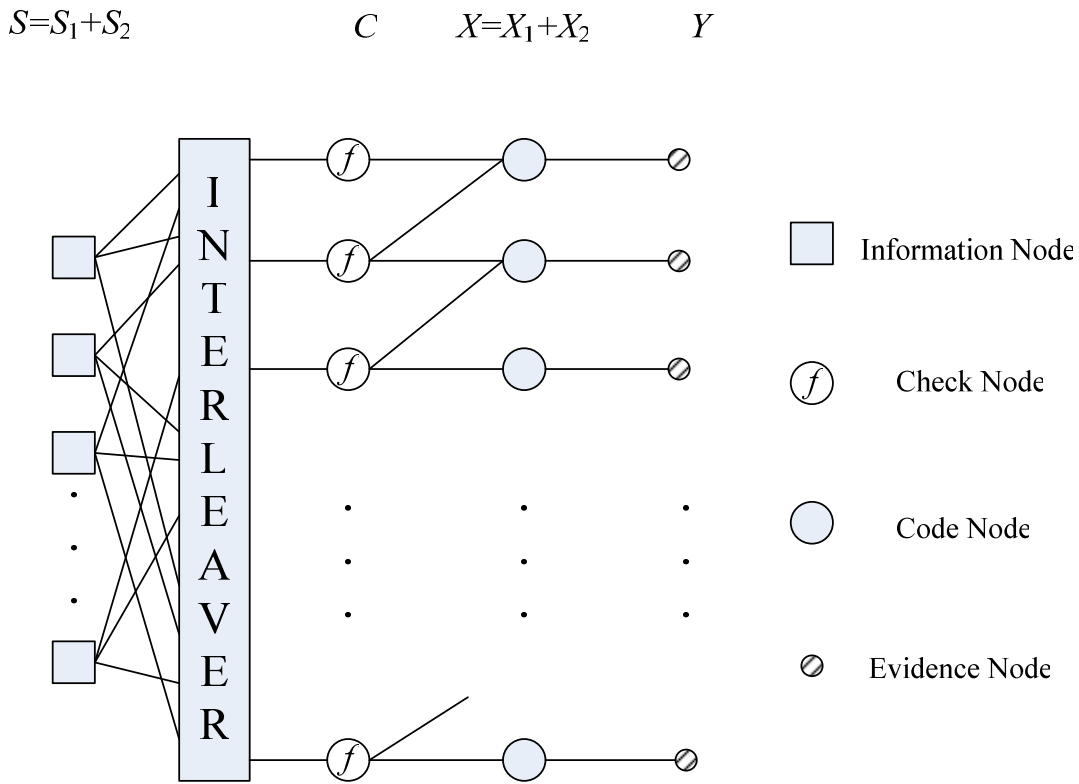

Fig 7:　Tanner graph of the virtual RA code in ACNC.

Generally speaking, the encoding and decoding operation based on the Tanner graph is as follows. For encoding, the Tanner graph is read from left to right. That is, symbols are passed from left to right. For decoding, the Tanner graph could be read backward from right to left. If there were no noise, given $(x_1+x_2)[k]$ for all $k$ received at the evidence nodes, $(s_1+s_2)[j]$ for all $j$ could be recovered at the information nodes in one iteration of message passing from right to left. The messages (a message is associated with one directional edge in the Tanner Graph) may simply contain the exact values of the symbol $(x_1+x_2)[k]$ or $(s_1+s_2)[j]$. With noise, instead of passing the symbol value from one node to the next, the a posteriori probabilities associated with the values are passed. Multiple iterations of message passing from right to left as in Fig. 8(a) and Fig. 8(b), and then from left to right as in Fig. 8(c) and Fig. 8(d), are needed [16, 17]. The idea is that after several iterations, the probabilities will converge and we could decode $(s_1+s_2)[j]$ based on them.

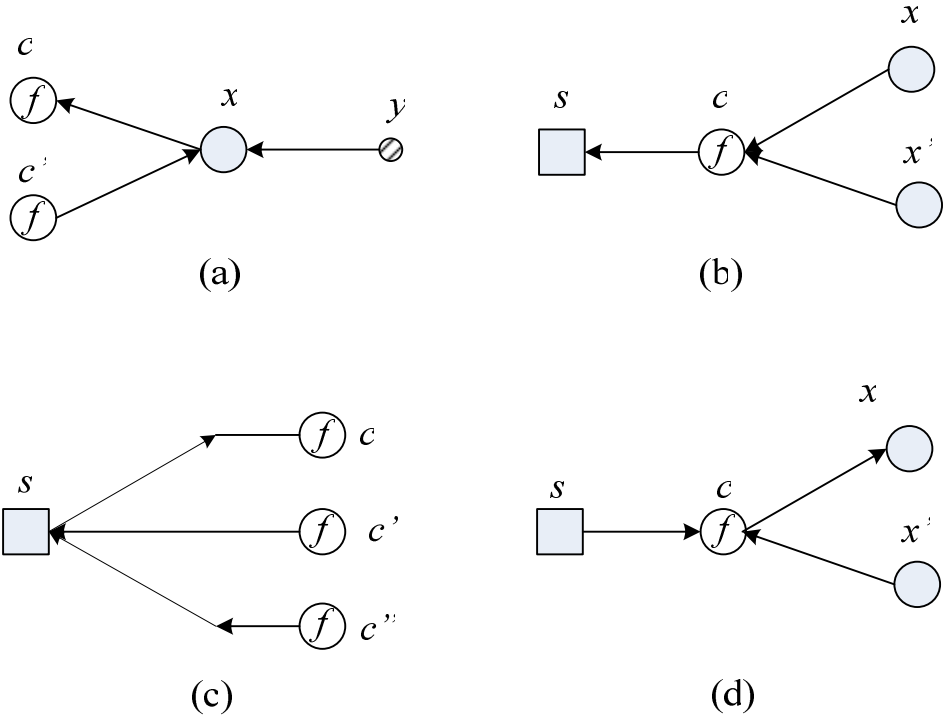

Fig 8: Message updating steps in one round of iteration.

*Step 3: Decoding algorithm*

With the Tanner graph in Fig. 7, we can design the particular decoding algorithm of the virtual encoder using a message passing mechanism similar to the generic message passing mechanism in [16]. The message form and the message update rules specific to our system are specified below.

We first rewrite the $k$-th received symbol at $N_3$ in (2) as

$$y_3[k] = (1 - 2x_1[k]) + (1 - 2x_2[k]) + n_3 \qquad (11)$$

The following algorithm can be extended to the case of general modulation as long as the received $q$-ary signal can be decomposed into $\log_2 q$ bits.

Let $P[h, t]$ denote the message passed between a check node and a variable node (information node or code node). The message is associated with the edge from node $h$ to node $t$, where one of $h$ or $t$ is a variable node, and the other is a check node. Let $P_k$, $k \in [1, qN]$, be the message from the $k$-th (ordered from top to bottom as in Fig. 7) evidence node to the $k$-th code node, where $N$ is the length of the uncoded packet.

*Message form:*

$P[h,t] = (p_0, p_1, p_2)$ is a vector, in which $p_i$ is the probability that the corresponding variable node ($h$ or $t$) takes on the value of $i$.

$P_k = (p_0, p_1, p_2)$ is a vector, in which $p_i$ is the probability that the $k$th coded symbol is $i$ given the $k$-th received symbol.

*Message Initial Values*:

All the messages associated with the edges in Fig. 7 are set to (1/4, 1/2, 1/4) except for the messages on the edges incident to the evidence nodes, which contain information on the received signal. The message from the evidence node $k$ is computed from the received signal $y_3[k]$ as follows:

$$\begin{aligned} P_k &= (p_0, p_1, p_2) \\ &= \left( \Pr((x_1+x_2)[k]=0 \mid y_3[k]), \Pr((x_1+x_2)[k]=1 \mid y_3[k]), \Pr((x_1+x_2)[k]=2 \mid y_3[k]) \right) \\ &= \frac{1}{\beta} \left( \exp(\frac{-(y_3[k]-2)^2}{2\sigma^2}), 2\exp(\frac{-(y_3[k])^2}{2\sigma^2}), \exp(\frac{-(y_3[k]+2)^2}{2\sigma^2}) \right) \end{aligned} \quad (12)$$

where $\beta$ is a normalizing factor given by $\beta = \exp(\frac{-(y_3[k])^2}{2\sigma^2}) \left( \exp(\frac{2y_3[k]-2}{\sigma^2}) + \exp(\frac{-2y_3[k]-2}{\sigma^2}) + 2 \right)$.

*Message Update Rules*:

Parallel to the generic updating rules in [16], we also have the same message updating rules at our check nodes and variable nodes. Note that the messages from the evidence nodes to the code nodes remain the same without being changed during the iterations of the decoding process.

*Update Equations for Output Messages Going Out of a Variable Node*

This is the case for Fig. 8(a) and (c). In the following, we focus on the scenario of Fig. 8(a). The update equations for the scenario of Fig. 8(c) are similar except that the variable node is an information node rather than a code node, and the associated probabilities are related to the source symbol rather than the code symbol. When the probability vectors of the two input messages, $P = (p_0, p_1, p_2)$ and $Q = (q_0, q_1, q_2)$ (associated with the edge from $y$ to $x$ and the edge from $c'$ to $x$, respectively), arrive at a code node of degree three (except the lowest code node), the probability that the code symbol is 0 is obtained as follows:

$$\begin{aligned}\Pr(x=0\,|\,P,Q) &= \frac{\Pr(P,Q\,|\,x=0)\Pr(x=0)}{\Pr(P,Q)} \\ &= \frac{\Pr(P\,|\,Q,x=0)\Pr(Q\,|\,x=0)\Pr(x=0)}{\Pr(P,Q)} \\ &= \frac{\Pr(P\,|\,x=0)\Pr(Q\,|\,x=0)\Pr(x=0)}{\Pr(P,Q)} \\ &= \frac{\Pr(x=0\,|\,P)\Pr(x=0\,|\,Q)\Pr(P)\Pr(Q)}{\Pr(P,Q)\Pr(x=0)} \\ &= 4\beta p_0 q_0\end{aligned} \qquad (13)$$

where $\beta = \dfrac{\Pr(P)\Pr(Q)}{\Pr(P,Q)}$ and the two input messages are assumed to be independent given the value of the variable node, i.e., $\Pr(P\,|\,Q,x) = \Pr(P\,|\,x)$. Given the $l$-depth neighborhood of the edge is cycle free (cycle free condition), this assumption is true for iterations up to $l$ in the decoding algorithm. As in the proof for the LDPC codes in [18], the probability that the cycle free condition is true for our coder in Fig. 7 should also go to 1 as the length of the code goes to infinity. That is, $l$ becomes larger and larger.

In a similar way, we can obtain that $\Pr(x=1\,|\,P,Q) = 2\beta p_1 q_1$ and $\Pr(x=2\,|\,P,Q) = 4\beta p_2 q_2$. Thus, the output message at the variable node is

$$VAR(P,Q) = 4\beta(p_0 q_0, p_1 q_1/2, p_2 q_2) \qquad (14)$$

Since the summation of the three probabilities should be 1, we require $\beta = (p_0 q_0 + p_1 q_1/2 + p_2 q_2)/4$ for normalization.

For the lowest code node in Fig. 7, the output message is always the same at the input message from the last evidence node, which remains constant throughout the iterations.

*Update Equations for Output Messages Going Out of Check Nodes:*

This is the case for Fig. 8(b) and (d) except that the accumulate function is $f$ in (10) instead of $\oplus$. We focus on the scenario of Fig. 8(b) here. Consider a check node below the topmost check node. Based on the $f$ defined in (10), and using similar computation as in (13), the probability that the information node symbol is 0 given the two input messages $P=(p_0,p_1,p_2)$ and $Q=(q_0,q_1,q_2)$ (associated with the edge from $x$ to $c$ and the edge from $x'$ to $c$, respectively) is

$$\begin{aligned}&\Pr(s=0|P,Q)\\&=\Pr(x=0,x'=0|P,Q)+\Pr(x=2,x'=2|P,Q)+\frac{1}{2}\Pr(x=1,x'=1|P,Q)\\&=\Pr(x=0|P)\Pr(x'=0|Q)+\Pr(x=2|P)\Pr(x'=2|Q)+\frac{1}{2}\Pr(x=1|P)\Pr(x'=1|Q)\\&=p_0q_0+p_2q_2+\frac{1}{2}p_1q_1\end{aligned} \quad (15)$$

In a similar way, we can obtain that $\Pr(s=1|P,Q)$ and $\Pr(s=2|P,Q)$. As a result, the output message at the check node is

$$CHK(P,Q)=(p_0q_0+p_1q_1/2+p_2q_2, p_1q_2+p_2q_1+p_1q_0+p_0q_1, p_0q_2+p_1q_1/2+p_2q_0) \quad (16)$$

For the topmost check node in Fig. 7, the output message is always the same at the input message from the topmost code node.

Notable is the fact that the complexity of our updating rules in (14) and (16) is indeed just four real-number multiplications ($p_0q_0, p_1q_1, p_0q_1$ and $p_1q_0$, others can be obtained with simple addition), which is same as the complexity of traditional RA decoder when the same message format is adopted. With the rules given in (14) and (16) and the initial message values given in (12), the detailed iterative belief propagation algorithm can be easily constructed as follows:

1. Set all the messages to the initial state.
2. Update messages iteratively as follows (i, ii, iii, and iv below corresponds to the settings in Fig. 8(a), (b), (c), and (d), respectively) :
   i. Update messages $P[x, c]$ and $P[x,c']$ at all code nodes $x \in X$, where $c$ and $c'$ are neighbor check nodes to $x$:

   If $x$ is the last code node at the bottom of Tanner graph,

   $P[x,c]=P_{qN}$

   If $x$ is not the last code node,

$$P[x,c] = VAR(P_k, P[c',x])$$
$$P[x,c'] = VAR(P_k, P[c,x])$$

    ii. Update messages $P[c, s]$ at all check nodes $c \in C$, where $s$, $x$, and $x'$ are neighbor variable nodes to $c$:

If $c$ is the first check node at the top of Tanner graph,

$$P[c,s] = P[x,c]$$

If $c$ is not the first check node at the top

$$P[c,s] = CHK(P[x,c], P[x',c])$$

    iii. Update messages $P[s,c], P[s,c'], P[s,c'']$ at all information nodes $s \in S$, where $c$, $c'$, and $c''$ are neighbor check nodes to $s$:

$$P[s,c] = VAR(P[c',s], P[c'',s])$$
$$P[s,c'] = VAR(P[c,s], P[c'',s])$$
$$P[s,c''] = VAR(P[c',s], P[c,s])$$

    iv. Update messages $P[c, x]$ at all check nodes $c \in C$, where $s$, $x$, and $x'$ are neighbor variable nodes to $c$:

If $c$ is the first check node at the top of Tanner graph,

$$P[c,x] = P[s,c]$$

If $c$ is not the first check node,

$$P[c,x] = CHK(P[s,c], P[x',c])$$
$$P[c,x'] = CHK(P[s,c], P[x,c])$$

    v. Go to step i until some criteria satisfied

3. When iteration stops, the output message for an information node $s$ is given by

$$P^v = VAR(VAR(P[s,c], P[s,c']), P[s,c''])$$

## V. Numerical Simulation

In this section, we investigate the performance of ACNC with the above decoding algorithm via numerical simulation. We set the repeat factor $q$ to 3 and the interleave pattern is randomly selected for each packet, but identical for all the three schemes. We apply ACNC and check the BER (bit error rate) of the decoded packet $S_1 \oplus S_2$ at the relay node. BPSK modulation is used at both end nodes and the power is equally allocated to them. The noise is AWGN with variance $\sigma^2$ and the SNR is defined as $1/\sigma^2$ (the total transmit power of the two end nodes is 2 and the average power of each one is 1).

For comparison, we also study the performance of CNC1 and CNC2 that use standard RA code. They use the same encoder as in ACNC, but the decoders at the relays are different. In CNC1, the two end nodes apply optimal power allocation as in eqn. (A-4) in [19]. The relay node obtains $P_{s_1,s_2}$ by successively decoding $Y_3$ to $S_1$ and $S_2$ with the standard SISO RA decoder sequentially and then combins them with (5). In CNC2, the relay $N_3$ transforms each symbol in $y_3$ to $P_{x_1 \oplus x_2}$ with the MMSE estimation as in [14] and then channel-decodes $X_1 \oplus X_2$ to $S_1 \oplus S_2$ using the standard RA decoder.

In Fig. 9, we show the BER performance of the three schemes under different SNR. In the simulation, the uncoded packet length is set to 4096 bits and the BER is calculated by averaging over 10,000 packets. The iteration numbers for both our new decoding algorithm and the standard RA decoding algorithm are set to 20, 30 or 40. As shown in Fig. 9, the BER of all three schemes decreases with the increase in SNR and the iteration number. ACNC outperforms CNC2 by about 0.5dB when the BER is in the ballpark of $10^{-4}$; and it outperforms CNC1 by an even larger gap. ACNC with 20-iteration decoding outperforms both CNC1 and CNC2 with 40-iteration decoding.

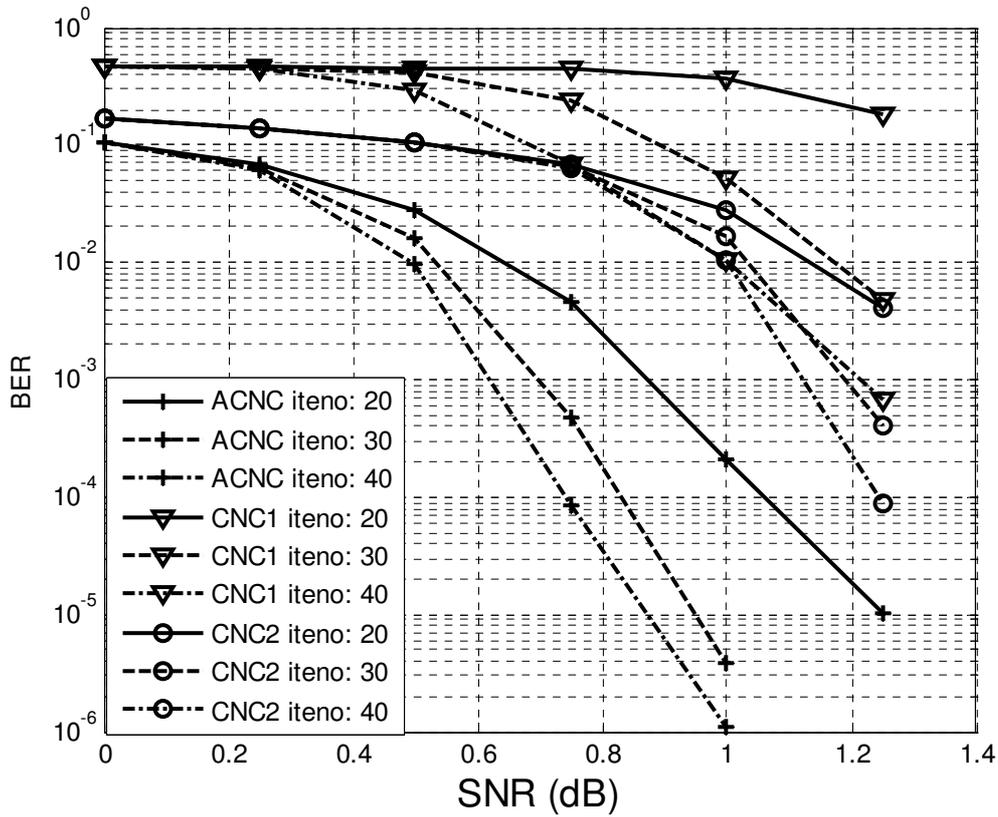

Fig 9: BER performance of the CNC1, CNC2, and ACNC, under different numbers of iterations used in the belief propagation algorithm.

In Fig. 10, we show the BER performance for different packet lengths (1024, 4096, and 8192 bits) when the iteration numbers of all three schemes are set to 30. In general, larger packet length leads to smaller BER for all the schemes. Fig. 10 also shows that for all packet lengths, we continue to observe the outperformance of ACNC over CNC2 by about 0.5 dB when the BER is $10^{-4}$; and the outperformance of ACNC over CNC1 by an even larger gap.

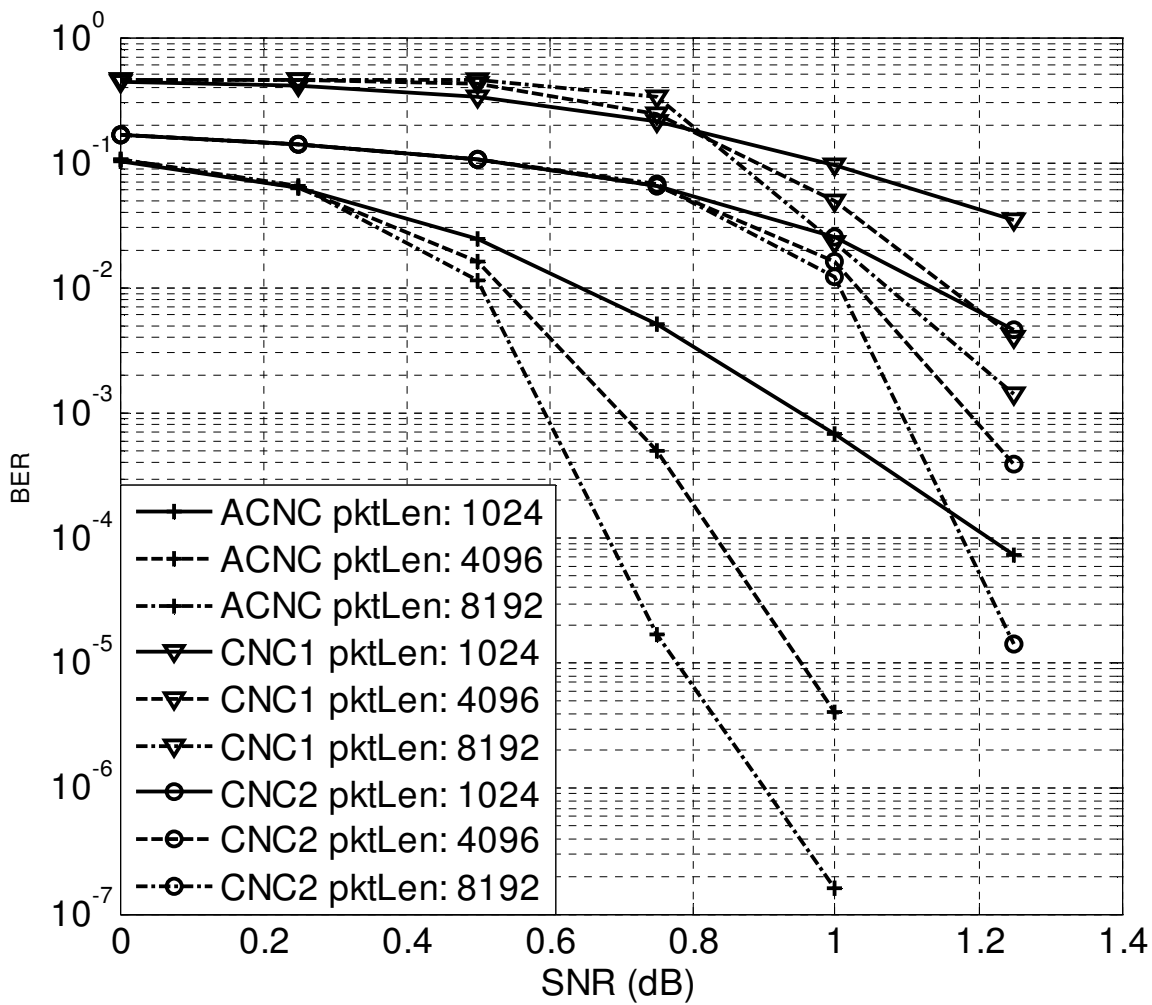

Fig 10: BER performance of the CNC1, CNC2, and ACNC, for various packet lengths.

## VI. Conclusion and Discussion

We have investigated three schemes for link-by-link coded PNC. The relative performance of the three schemes lies in the Channel-decoding-Network-Coding (CNC) strategies used at the relay node. In particular, an insight from this paper is that we should (i) avoid decoding extraneous information not related to $S_1 \oplus S_2$; (ii) make full use of the information contained in $Y_3$ to help decode the network-coded packets $S_1 \oplus S_2$. Guided by these two principles, an Arithmetic-sum CNC (ACNC) scheme has been proposed in this paper. Specifically, we provide an implementation of ACNC based on RA code and a special belief propagation decoding algorithm tailored for PNC mapping.

For comparison purposes, two conventional CNC schemes, CNC1 and CNC2, have been investigated. From viewpoint of the two design principles, our ACNC scheme avoids the shortcomings of CNC1 and CNC2 while preserving the advantages of them without added decoding complexity. Our simulation indicates that ACNC can have substantial BER improvements over CNC1 and CNC2.

In [8-11], it was proved that CNC1 and CNC2 can reliably transmit $S_1 \oplus S_2$ to the relay with a rate approaching the capacity in low and high SNR regions, respectively. Since our investigation indicates that ACNC can outperform both CNC1 and CNC2 when the RA code is used, we conjecture that ACNC by itself could approach the capacity of TWRC in both low and high SNR regions. In the appendix of our technical report [19], we derive a prospective rate of ACNC. The prospective rate, which is higher than the rates of both CNC1 and CNC2 for all SNR, provides an intuition as to the plausibility of our conjecture. A rigorous proof, however, awaits further investigation.

**Appendix I: Decoding algorithm with non-perfect synchronization**

The proposed joint decoding algorithm in section IV is based on the assumption that perfect synchronization is achieved between the two end nodes, i.e. the signals from $N_1$ and $N_2$ arrive at the relay node with the same power, the same phase, and at the same time. In practice, however, it is difficult to achieve such perfect synchronization, especially in a fading channel. As shown in [5], the non-perfect synchronization will result in power penalties, and we can express this effect with fading coefficients $\sqrt{P_1}, \sqrt{P_2}$. Then the received signal the relay node is the same as in (2). We now discuss the joint decoding algorithm when $P_1 \neq P_2$, $P_1 + P_2 = 2$.

The first way is to keep the virtual encoder in section IV unchanged. Then, its output is $x_v[k] = x_1[k] + x_2[k] = f(x_v[k-1], u_v[k])$ and the Gaussian noise and unequal power allocation in the received packet $Y_3 = \sqrt{P_1} X_1 + \sqrt{P_2} X_2 + W_3$ is regarded as the effect of the channel. Then the decoding algorithm is identical to the one in section IV except that the initial message value in (12) needs to be changed to

$$\begin{aligned} P_k &= (p_0, p_1, p_2) h \\ &= \left( \Pr((x_1+x_2)[k]=0 \mid y_3[k]), \Pr((x_1+x_2)[k]=1 \mid y_3[k]), \Pr((x_1+x_2)[k]=2 \mid y_3[k]) \right) \\ &= \frac{1}{\beta} \left( \exp(\frac{-(y_3[k]-\sqrt{P_1}-\sqrt{P_2})^2}{2\sigma^2}), \right. \\ &\quad \exp(\frac{-(y_3[k]-\sqrt{P_1}+\sqrt{P_2})^2}{2\sigma^2}) + \exp(\frac{-(y_3[k]-\sqrt{P_2}+\sqrt{P_1})^2}{2\sigma^2}), \\ &\quad \left. \exp(\frac{-(y_3[k]+\sqrt{P_1}+\sqrt{P_2})^2}{2\sigma^2}) \right) \end{aligned}$$

The other way is to construct a new function $g$ instead of $f$ such that the output of the virtual encoder is $x_v[k] = \sqrt{P_1} x_1[k] + \sqrt{P_2} x_2[k] = g(x_v[k-1], u_v[k])$. Similar to (10), we can obtain the exact formulation of function $g$ as

$$x_v[k] = g(x_v[k-1], u_v[k]) = \sqrt{P_1}x_1[k-1] \oplus s_1[j] + \sqrt{P_2}x_2[k-1] \oplus s_2[j]$$

$$= \begin{cases} 0 & \text{if } x_v[k-1] = \gamma_3, (s_1+s_2)[j] = 2 \\ \gamma_4 & \text{if } x_v[k-1] = \gamma_3, (s_1+s_2)[j] = 1 \\ \gamma_3 & \text{if } x_v[k-1] = \gamma_3, (s_1+s_2)[j] = 0 \\ \gamma_2 & \text{if } x_v[k-1] = \gamma_1, (s_1+s_2)[j] = 2 \\ \gamma_1 & \text{if } x_v[k-1] = \gamma_2, (s_1+s_2)[j] = 2 \\ 0 \text{ or } \gamma_3 & \text{if } x_v[k-1] = \gamma_4, (s_1+s_2)[j] = 1 \\ \gamma_1 & \text{if } x_v[k-1] = \gamma_1, (s_1+s_2)[j] = 0 \\ \gamma_2 & \text{if } x_v[k-1] = \gamma_2, (s_1+s_2)[j] = 0 \\ \gamma_3 & \text{if } x_v[k-1] = 0, (s_1+s_2)[j] = 2 \\ \gamma_4 & \text{if } x_v[k-1] = 0, (s_1+s_2)[j] = 1 \\ 0 & \text{if } x_v[k-1] = 0, (s_1+s_2)[j] = 0 \end{cases}$$

where $\gamma_1 = \sqrt{P_1}, \gamma_2 = \sqrt{P_2}, \gamma_3 = \sqrt{P_1} + \sqrt{P_2}, \gamma_4 = \sqrt{P_1} \text{ or } \sqrt{P_2}$. It is not difficult to find that the function $g$ satisfies the property (a) in section IV. Note that there are two possibilities for $g(x_v[k], x_v[k-1])$. The first is $g(x_v[k], x_v[k-1]) = 0$ and the second is $g(x_v[k], x_v[k-1]) = P_1 + P_2 = 2$. The third is $g(x_v[k], x_v[k-1]) = P_1$ or $P_2$ (with equal probability), in which case $g(x_v[k], x_v[k-1])$ will be mapped to 1. Then we can find that the function $g$ also satisfies the property (b) in section IV. , With the function $g$, we can design the updating rules in a similar way as in (14) and (16) to obtain the new decoding algorithm.